\documentclass{llncs}

\usepackage{amssymb}
\usepackage{amsmath}
\usepackage{amsfonts} 
\usepackage[ruled, vlined, linesnumbered, commentsnumbered]{algorithm2e} 
\usepackage{tabularx}
\usepackage{tabulary}
\usepackage{pdflscape}
\usepackage{color}

\begin{document}

\title{Accurate Decoding of Pooled Sequenced Data Using Compressed Sensing}

\author{Denisa~Duma\inst{1} \and Mary~Wootters\inst{2} \and 
Anna~C.~Gilbert\inst{2}  \and \\ Hung~Q.~Ngo\inst{3} \and  Atri~Rudra\inst{3} \and 
Matthew~Alpert\inst{1}  \and \\ Timothy~J.~Close\inst{4} \and Gianfranco~Ciardo\inst{1}  \and 
Stefano~Lonardi\inst{1}}

\institute{Dept. of Computer Science and Eng., University of California, Riverside, CA~92521 \\
\and
Dept. of Mathematics, The University of Michigan, Ann Arbor, MI~48109 \\
\and 
Dept. of Computer Science and Eng., Univ. at Buffalo, SUNY, Buffalo, NY~14260 \\
\and
Dept. of Botany \& Plant Sciences, University of California, Riverside, CA~92521}

\maketitle  

\begin{abstract}
In order to overcome the limitations imposed by DNA barcoding when
multiplexing a large number of samples in the current generation of high-throughput sequencing
instruments, we have recently proposed a new protocol that leverages
advances in combinatorial pooling design (group testing) \cite{10.1371/journal.pcbi.1003010}. We have also demonstrated how this new protocol would enable \emph{de novo} selective sequencing and assembly of large, highly-repetitive
genomes.  Here we address the problem of decoding pooled sequenced
data obtained from such a protocol.  Our algorithm employs a
synergistic combination of ideas from compressed sensing and the
decoding of error-correcting codes. Experimental results on synthetic
data for the rice genome and real data for the barley genome show that
our novel decoding algorithm enables significantly higher quality
assemblies than the previous approach.

  
\end{abstract}


\section{Introduction}

The second generation of DNA sequencing instruments offer
unprecedented throu\-ghput and extremely low cost per base, but read
lengths are much shorter compared to Sanger sequencing. An additional
limitation is the small number of distinct samples that these
instruments can accommodate (e.g., two sets of eight lanes on the
Illumina HiSeq).  When the sequencing task involves a large number of
individual samples, a common solution is to employ DNA barcoding
to ``multiplex'' samples within a single lane.  DNA barcoding,
however, does not scale readily to thousands of samples. As the number
of samples reaches the hundreds, exhaustive DNA barcoding becomes
time consuming, error-prone, and expensive.  Additionally, the
resulting distribution of reads for each barcoded sample can be
severely skewed (see, e.g., \cite{BarcodeBias}).

\emph{Combinatorial pooling design} or \emph{group testing} allows
one to achieve multiplexing without exhaustive barcoding. In group
testing, a \emph{design} or \emph{scheme} is a set of tests (or pools)
each of which is a subset of a large collection of items
that needs to be tested for the presence of (a few) `defective' items.
The result of testing a pool is a boolean value indicating whether
the pool contains at least one defective. The goal of group testing is to
\emph{decode} the information obtained by testing all the pools in order to
determine the precise identity of the defectives, despite the fact that the
defectives and non-defectives are mixed together. The challenge is to achieve
this goal while, at the same time, minimizing the number of pools needed.
Recently, \emph{compressed sensing} (CS) has emerged as a powerful technique
for solving the decoding problem when the results of testing the pools
are more than boolean outcomes, for instance, real or complex
values. 

Combinatorial pooling has been used previously in the context of
genome analysis (see, e.g.,
\cite{YanivErlich072009,Erlich:2010je,Hajirasouliha:2008,Amir&Zuk,Shental:2010bu}),
but not for \emph{de novo} genome sequencing. Our proposed pooling
method for genome sequencing and assembly was first described in \cite{10.1371/journal.pcbi.1003010}
and has generated considerable attention. 
It was used to produce one of the critical datasets for the first draft sequence of the barley genome
\cite{Nature12}. In our sequencing protocol, thousands of BAC
clones are pooled according to a combinatorial
design so that, at the outset of sequencing, one can `decode' each
read to its source BACs. The underlying idea is to encode the
identity of a BAC within the pooling pattern rather than by its
association with a specific DNA barcode.  We should stress that
combinatorial pooling is not necessarily an alternative to DNA
barcoding, and both methods have advantages and disadvantages. 
They can be used together to increase the number of
samples that can be handled and benefit from the advantages of both.

In this paper we address the problem of decoding pooled sequenced
data obtained from a protocol such as the one in \cite{10.1371/journal.pcbi.1003010}.
While the main objective is to achieve the highest possible accuracy
in assigning a read to the correct BAC, given that one sequencing run
can generate hundreds of millions of reads, the
decoding procedure has also to be time- and space-efficient.  Since
in \cite{10.1371/journal.pcbi.1003010} we pooled BAC clones according to the \emph{Shifted Transversal Design}
\cite{Thierry2006a} which is a \emph{Reed-Solomon} based pooling design, our
proposed decoding approach combines ideas from the fields of
compressive sensing and decoding of error-correcting codes. Specifically, given the result of `testing'
(in this case, sequencing) pools of genomic BAC clones, we aggregate
read frequency information across the pools and cast the problem as a compressed
sensing problem where the unknowns are the BAC assignments of the reads. We
solve (decode) for the unknown assignments using a \emph{list recovery} strategy as
used in the decoding of error-correcting codes. Reed-Solomon codes are known to be good list-recoverable codes which can also tolerate a large fraction of errors. We also show that using readily available information about the reads
like overlap and mate pair information can improve the accuracy of the decoding.
Experimental results on synthetic reads from the rice genome as well as real sequencing reads from the barley genome show that the decoding accuracy of our new method is almost identical to that of
\textsc{HashFilter} \cite{10.1371/journal.pcbi.1003010}.  However,
when the assembly quality of individual BAC clones is the metric of
choice, the decoding accuracy of the method proposed here is
significantly better than \textsc{HashFilter}.



\section{Related Work}
\label{sec:related}

The resemblance between our work and the closest related research
efforts using combinatorial pooling and compressed sensing ideas stops at the pooling of sequencing data. Our application domain, pooling scheme employed and algorithmic approach to decoding, are
completely different. To the best of our knowledge, all compressed sensing
work in the domain of genomics deals with the problem of \emph{genotyping} large population
samples, whereas our work deals with \emph{de novo} genome sequencing. For instance in
\cite{YanivErlich072009}, the authors employ a pooling scheme based on the Chinese
Remainder Theorem (CRT) to identify carriers of rare alleles in large cohorts
of individuals.  The pooling scheme allows the detection of
mutants within a pool, and by combining information across pools one is able to
determine the identity of carriers. In true group testing style, the unknown carrier identities are encoded
by a boolean vector of length equal to the number of individuals, where a
value of one indicates a carrier and zero a normal individual. To decode their pooling scheme and find the unknown vector, the authors devise a greedy decoding method called \emph{Minimum Discrepancy Decoder}.  In \cite{Erlich:2010je},
loopy belief propagation decoding is used for the same pooling scheme.
A similar application domain is described in \cite{Shental:2010bu},
where the authors identify carriers of rare SNPs in  a group of individuals pooled with a random
pooling scheme (Bernoulli matrix) and use the \emph{Gradient
  Projection for Sparse Reconstruction} (GPSR) algorithm to decode the
pooling scheme and recover the unknown carrier identities. The same problem is tackled in
\cite{SnehitPrabhu072009} with a pooling design inspired from the
theory of error correcting codes. However, this design is only able to
identify a single rare-allele carrier within a group. In \cite{Amir&Zuk}, the authors organize domain-specific (linear) constraints into a compressed sensing matrix which they use together with GPSR decoding to determine the frequency of each bacterial species present in a metagenomic mixture.


\section{Preliminaries}
\label{sec:preliminaries}

As mentioned in the introduction, in \cite{10.1371/journal.pcbi.1003010} we pool DNA samples (BAC
clones) according to a combinatorial pooling scheme, then sequence the
pools using high-throughput sequencing instruments. In this paper we show how to efficiently recover the sequence content of each BAC by combining ideas from the theory of
\emph{sparse signal recovery} or \emph{compressed sensing} (CS) as
well as from the large body of work developed for the decoding of \emph{error-correcting codes}. 

Formally, a combinatorial pooling design (or pooling scheme) can
be represented by a binary matrix $\mathbf{\Phi}$ with $m$
rows (corresponding to pools) and $n$ columns (corresponding to items
to be pooled), where entry $(i,j)$ is $1$ if item $j$ is present in
pool $i$, $0$ otherwise. The matrix $\mathbf{\Phi}$ is called the
\emph{design matrix}, \emph{sensing matrix} or \emph{measurement
  matrix} by various authors in the literature. In this paper we 
only use the first two names to designate $\mathbf{\Phi}$. An important property
of a combinatorial pooling design is its \emph{decodability} \emph{d} (also
called \emph{disjunctness}), which is the maximum
number of `defectives' it guarantees to reliably identify.  
Let $w$ be a subset of the columns (pooled
  variables) of the design matrix $\mathbf{\Phi}$ and $p(w)$ be the
  set of rows (pools) that contain at least one variable in
  $w$: the matrix $\mathbf{\Phi}$ is said to be $d$-decodable ($d$-disjunct) if
for any choice of $w_1$ and $w_2$ with $|w_1| = 1$, $|w_2| = d$ and $w_1 \not \subset w_2$,
we have that $p(w_1) \not \subseteq p(w_2)$.

In this paper, we pool BACs using the combinatorial pooling scheme called
\emph{Shifted Transversal Design} (STD) \cite{Thierry2006a}. STD is
a \emph{layered} design, \emph{i.e.}, the rows
of the design matrix are organized into multiple redundant layers
such that each pooled variable appears only once in each layer, that is, a \emph{layer} is a partition of the set of variables. STD is defined by
 parameters $(q, L, \Gamma)$ where $L$ is the number of layers,
$q$ is a prime number equal to the number of pools (rows) in each
layer and $\Gamma$ is the \emph{compression level} of the design.
Thus, in order to pool $n$ variables, STD uses a total of $m=q \times L$ pools.
The set of $L$ pools defines a unique pooling pattern for each variable which
can be used to retrieve its identity. This set of $L$ integers is called the
\emph{signature} of the variable. 
The compression level $\Gamma$ is defined to be the
smallest integer such that $q^{\Gamma+1} \ge n$.  STD has the desirable property that
any two variables co-occur in at most $\Gamma$ pools, therefore
by choosing a small value for $\Gamma$ one can make STD pooling
extremely robust to errors. 
The parameter $\Gamma$ is also related to
the decodability of the design through the equation
$d=\lfloor{(L-1)/\Gamma}\rfloor$. Therefore, $\Gamma$ can be seen as a trade-off parameter: the larger it is, the more items can be tested (up to $q^{\Gamma+1}$), but fewer defectives can be reliably identified (up
to $\lfloor(L-1)/\Gamma\rfloor$). For more details on the pooling scheme
and its properties please refer to \cite{Thierry2006a}.



In order to decode measurements obtained through STD (\emph{i.e.}, reconstruct the sequence content of pooled BACs) we borrow ideas from compressed sensing (CS), an area of signal
processing that describes conditions and efficient methods
for capturing sparse signals from a small number of aggregated
measurements \cite{Erlich:2010je}.
Unlike combinatorial group testing, in
compressed sensing measurements can be more general than boolean
values, allowing recovery of hidden variables which are
real or complex-valued. 
Specifically, in CS we look for an unknown vector or \emph{signal}
$\mathbf{x}=(x_1, x_2, \dots, x_n)$ which is \emph{$s$-sparse}, \emph{i.e.}, has at most $s$ non-zero
entries. We are given a vector $\mathbf{y}=(y_1, y_2, \dots, y_m)$ of
measurements $(m \ll n)$, which is the product between the (known) design
matrix $\mathbf{\Phi}$ and the unknown vector $\mathbf{x}$, that is
$\mathbf{y}=\mathbf{\Phi} \mathbf{x}$. Under certain conditions
on $\mathbf{\Phi}$, by using the measurements $\mathbf{y}$, the
assumption on the sparsity of $\mathbf{x}$ and
information encoded by $\mathbf{\Phi}$, it is possible to recover the
original sparse vector $\mathbf{x}$. The latter equation corresponds to
the ideal case when the data is noise-free. In practice, if
the signal $\mathbf{x}$ is not as sparse as needed and if 
measurements are corrupted by noise, the equation
becomes $\mathbf{y} = \mathbf{\Phi} \mathbf{x} + \epsilon$. 
In CS theory there are two main approaches for solving the latter
equation, namely \emph{linear programming} (LP) decoding and \emph{greedy pursuit}
decoding.
Greedy pursuit algorithms
have faster decoding time than LP-based approaches, frequently sub-linear in the length of $\mathbf{x}$ (although for specially designed matrices).
Their main disadvantages is that they usually require a
slightly larger number of measurements and do not offer the same uniformity and stability guarantees as LP decoding.
Greedy pursuits are iterative algorithms which proceed in a series of
steps: (1) identify the locations of the largest coefficients of
$\textbf{x}$ by greedy selection, (2) estimate their values, 
(3) update $\mathbf{y}$ by subtracting the
contribution of estimated values from it, and iterate (1-3) until some
convergence criterion is met. Usually $O(s)$ iterations, where $s$ is the
sparsity of $\mathbf{x}$, suffice \cite{SOMP}. 
Updating $\mathbf{y}$ amounts to solving a least squares problem in each
iteration.    

The most well known greedy decoding algorithm is \emph{Orthogonal
Matching Pursuit} (OMP) \cite{OMP}, which has spawned many
variations.  In OMP, the greedy rule selects in each iteration the
largest coordinate of $\mathbf{\Phi}^T\mathbf{y}$, \emph{i.e.}, the
column of $\mathbf{\Phi}$ which is the most correlated with
$\mathbf{y}$.  In this paper, we are interested in a
variant of OMP called \emph{Simultaneous Orthogonal
Matching Pursuit} (S-OMP). S-OMP is different from OMP in that
it approximates multiple sparse signals $\mathbf{x}_1, \mathbf{x}_2,
\dots, \mathbf{x}_K$ simultaneously by using multiple linear
combinations, $\mathbf{y}_1, \mathbf{y}_2, \dots, \mathbf{y}_K$,
of the sensing matrix $\mathbf{\Phi}$ \cite{SOMP}. The
unknown signals $\{\mathbf{x}_k\}_{k\in \{1,\cdots,K\}}$ as well as
measurement vectors $\{\mathbf{y}_k\}_{k \in \{1,\cdots, K\}}$
can be represented by matrices $\mathbf{X} \in \mathcal{R}^{n \times K}$
and $\mathbf{Y} \in \mathcal{R}^{m \times K}$.  Intuitively, by jointly
exploiting information provided by $\mathbf{Y}$, S-OMP is able to
achieve better approximation error especially when the signals to be
approximated are corrupted by noise which is not statistically
independent \cite{SOMP}.

The mapping from the CS setting into our problem follows naturally and
we give here a simplified and intuitive version of it.  The detailed
model will be introduced in the next section. The variables to be
pooled are BAC clones. Each column of the design matrix
corresponds to a BAC to be pooled and each row corresponds to a
pool. For each read $r$ (to be decoded) there is an unknown $s$-sparse
vector $\mathbf{x}$ which represents at most $s$ BACs which could have generated $r$. The vector of 
measurements $\mathbf{y}$ (\emph{frequency vector}) of length $m$ gives for
each read $r$, the number of times $r$ appears in each of the $m$
pools. The use of numerical measurements (read counts) rather than
boolean values indicating the presence or the absence of $r$ from a
pool is in accordance with CS theory and offers additional valuable
information for decoding.  To carry out the latter, we use a S-OMP
style algorithm but replace the greedy selection rule by a
\emph{list recovery} criterion.  Briefly, we obtain a list
of candidate BACs for read $r$ as those columns of $\mathbf{\Phi}$ whose non-zero
coordinates consistently correspond to the heaviest-magnitude
measurements in each layer of $\mathbf{y}$
\cite{NgoPoratRudra12}. This allows for a finer-grained usage of the
values of $\mathbf{y}$ on a layer-by-layer basis rather than as a
whole. Additionally, by requiring that the condition
holds for at least $\mathit{l}$ layers with $\mathit{l} \leq L$, one can make the algorithm more robust to the noise in vector $\mathbf{y}$.
  

\section{Decoding Algorithms}
\label{sec:algorithm}

In this section we present our decoding algorithms that assign reads
back to the BACs from which they were derived. Recall that we have $n$ BACs pooled into $m$
pools according to STD and each BAC is pooled in exactly
$L$ pools. The input data to the decoding algorithm consists of (1)
$m$ datasets containing the reads obtained from sequencing the $m$ pools,
and (2) the parameters of the pooling design, including the signatures
of all $n$ BACs. We will assume that each read
$r$ may originate from up to $s$ BACs with $s \ll n$;
ideally, we can make the same assumption for each
$k$-mer (a $k$-mer is a substring of $r$ of length $k$)
of $r$, provided that $k$ is `large enough'. In practice, this will not
be true for all $k$-mers (e.g., some $k$-mers are highly repetitive),
and we will address this issue later in this document.

We start by preprocessing the reads to correct sequencing errors in order to improve the accuracy of read decoding. For this task, we employ \textsc{SGA} \cite{SGA}, which internally employs a $k$-mer based
error correction strategy. An additional benefit of error
correction is that it reduces the total number of distinct $k$-mers present in the set of reads.
After the application of \textsc{SGA}, there still remains a small
proportion of erroneous $k$-mers, which we discard because
they will likely introduce noise in the decoding process.  An
advantage of pooled sequencing is that erroneous
$k$-mers are easy to identify because they appear in fewer
than $L$ pools. To be conservative, we only discard $k$-mers appearing
in fewer than $\gamma$ pools where $\gamma \leq L$ is a user-defined
parameter (see Section~\ref{sec:rice} for details on the choice of this parameter).
The closer $\gamma$ is to $L$ the more likely it is that a
$k$-mer that appears in $\gamma$ pools is correct, but
missing from the remaining $L-\gamma$ pools due to sequencing
errors. Henceforth, we will call a $k$-mer \emph{valid} if it
appears in a number of pools in the range $[\gamma,s L]$ where $s$ is the
sparsity parameter.  Any $k$-mer occurring in more than
$s L$ pools is considered highly repetitive, and will likely not be
useful in the decoding process. The decoding algorithm we employ
can safely ignore these repetitive $k$-mers.

To carry out the decoding, we first compute the frequencies of all the $k$-mers in all the $m$ pools. 
Specifically, we decompose all
\textsc{SGA}-corrected reads into $\emph{k}$-mers by sliding a
window of length $k$ (there are $|r|-k+1$ such windows for each read $r$). For each distinct $k$-mer, we
count the number of times it appears in each of the $m$ pools, and
store the sequence of the $k$-mer along with its vector of $m$ counts into a
hash table. We refer to the vector of counts of a
$k$-mer as its \emph{frequency vector}.

We are now ready to apply our CS-style decoding algorithm. 
We are given a large number of reads divided into $m$ sets (pools). For each read $r$, we want to
determine which of the $n$ BACs is the source. 
Since we decomposed $r$ into its constitutive $k$-mers, we
can represent the pool counts of all its $k$-mers by a \emph{frequency matrix}
$\mathbf{Y}_r$. Matrix $\mathbf{Y}_r$ is a non-negative integer
matrix where the number of columns is equal to the number $K_r$ of $k$-mers in
$r$, the number of rows is equal to the numbers $m$ of pools, and entry $(i,j)$ reports the number
of times the $j^{th}$ $k$-mer of $r$ appears in pool $i$.
The input to the decoding algorithm for read $r$ is given by (1) the frequency matrix
$\mathbf{Y}_r$, (2) the design matrix $\mathbf{\Phi} \in
\{0,1\}^{m \times n}$, and (3) the maximum number $s$ of BACs which could
have generated $r$. To decode $r$ means to find a matrix
  $\mathbf{X}_r \in \mathbf{Z}^{n \times K_r}$ such that $\mathbf{X}_r
  = \rm{argmin}_{\mathbf{X}}{||\mathbf{\Phi} \mathbf{X} - \mathbf{Y}_r||_2}$
  with the constrain that $\mathbf{X}_r$ is row-sparse, \emph{i.e.,} it has
  at most $s$ non-zero rows (one for each source BAC). 

Since finding the source BACs for a read is sufficient for our
purposes, we can reduce the problem of finding matrix
$\mathbf{X}$ to the problem of finding its \emph{row support} $S(\mathbf{X})$, which
is the union of the supports of its columns. The support
$\it{Supp}(\mathbf{X}_{:,j})$ of a column $j$ of $\mathbf{X}$ is the
set of indices $i$ such that $\mathbf{X}_{i,j} \neq 0$. In our case,
the non-zero indices represent the set of BACs which generated the
read (and by transitivity its constitutive $k$-mers). Since
this set has cardinality at most $s$, in the ideal case, $\mathbf{X}$
is row-sparse with support size at most $s$.  In practice, the same
$k$-mer can be shared by multiple reads and therefore the number of
non-zero indices can differ from $s$. By taking a conservative
approach, we search for a \emph{good} $s$-sparse approximation of
$S(\mathbf{X})$, whose quality we evaluate according to the following
definition.

\textbf{Definition:} A non-empty set $S$ is \emph{good} for $\mathbf{X}$ if
for any column $j$ of $\mathbf{X}$, we have $S \subset
\it{Supp}(\mathbf{X}_{:,j})$.


Our decoding Algorithm~\ref{algo:findsupport} finds $S$ in two steps,
namely \textsc{Filter} and \textsc{Estimate}, which are explained next.

\begin{algorithm}[b]
	\DontPrintSemicolon
	\SetFuncSty{textsc}
	\SetAlCapNameFnt{\sc}
	\SetKwData{Left}{left}\SetKwData{This}{this}\SetKwData{Up}{up}
	\SetKwFunction{Filter}{Filter}\SetKwFunction{Estimate}{Estimate}
	\SetKwInOut{Input}{Input}\SetKwInOut{Output}{Output}
	\Input {$\mathbf{\Phi} \in \{0,1\}^{m \times n}$, $\mathbf{Y}_r \in \mathbf{N}^{m \times K_r}$ and sparsity $s$ such that $\mathbf{X}_r = argmin_{\mathbf{X}}{||\mathbf{\Phi} \mathbf{X} - \mathbf{Y}_r||_2}$ for a $s$-row-sparse matrix $\mathbf{X}_r \in \mathbf{N}^{n \times K_r}$; $h \leq q$ the number of entries per layer considered by list recovery} 
	
	\Output {A non-empty set $S_r$ with $|S_r| \leq s$ which is \emph{good} for $\mathbf{X}_r$}
	\BlankLine
	$\mathbf{\tilde{X}_r} \leftarrow $ \Filter{$\mathbf{\Phi}, \mathbf{Y}_r, h$}\;
	$S_r \leftarrow$ \Estimate{$\mathbf{\tilde{X}}_r, s$}\;
	\Return{$S_r$} 
	\caption{FindSupport $(\mathbf{\Phi},\mathbf{Y}_r,h,s)$}\label{algo:findsupport}
\end{algorithm}

\textsc{Filter} (Algorithm~\ref{algo:filter}) is a one-iteration S-OMP style algorithm in
which multiple candidate BACs are selected (we tried performing
multiple iterations without significant improvement in the results).
Whereas S-OMP selects one BAC per iteration as the column of $\mathbf{\Phi}$ most correlated (inner product) with all the columns of $\mathbf{Y}$, our algorithm employs a list recovery criterion to obtain an
approximation $\mathbf{\tilde{X}}_r$ of $\mathbf{X}_r$. 
Specifically, for each column $y$ of $\mathbf{Y}_r$ and for each
layer $l \in [1,\dots,L]$, we select a set $S_l$ of candidate
pools for that layer as follows. We choose set $S_l$ by considering the
$h$ highest-magnitude coordinates of $y$ in layer $l$ and selecting
the corresponding pools. BACs whose signature pools
belong to all $L$ sets $S_l$ are kept while the rest of them
are removed, \emph{i.e.}, their $\tilde{\mathbf X}$-entries are set to zero.
Finally, for the BACs that are not filtered out, the
$\tilde{\mathbf X}$-entry estimate follows the \emph{min-count} 
 estimate. The value of $h$ should be chosen to be
$\Theta(s)$: $h=3s$ is sufficient even for noisy data \cite{NgoPoratRudra12}.     

\begin{algorithm}[t]
	\DontPrintSemicolon
	\SetFuncSty{textsc}
	\SetAlCapNameFnt{\sc}
	\SetKwData{lm}{layersMatched}
	\SetKwInOut{Input}{Input}
	\SetKwInOut{Output}{Output}

	\Input {$\mathbf{\Phi} \in \{0,1\}^{m \times n}, \mathbf{Y}_r \in \mathbf{N}^{m \times K_r}$, parameter $h$}
	\Output {An approximation $\mathbf{\tilde{X}}_r$ for $\mathbf{X}_r$}
	\BlankLine
	\tcp{Recall that $\mathbf{\Phi}$ has $L$ layers with $q$ pools each} 
	\tcp{For a column $y$ of $\mathbf{Y}_r$, denote by $y[l]_i$ the $i^{th}$ entry in layer $l$} 
	
	\BlankLine

	$\mathbf{\tilde{X}}_r \leftarrow 0$\;
	\For{$k = 1, \dots, K_r$} {
	Let $y=\mathbf{Y}_{r_{:,k}}$ be the $k^{th}$ column of $\mathbf{Y}_r$\;
		\For { $l = 1, \dots, L$}{
			$S_l \leftarrow$  set of $h$ indices $i \in \{1, \dots, q\}$ such that the corresponding counts
			$y[l]_{i}$ are the $h$ heaviest-magnitude counts in layer $l$ of column $y$\;
		}
			 \For{$b=1, \dots, n$}{
			 	$\lm \leftarrow 0$\;
				Let $\phi=\mathbf{\Phi_{:,b}}$ be the $b^{th}$ column of $\mathbf{\Phi}$\;
				\For { $l = 1, \dots, L$} {
					\If {the unique $i$ such that $\phi[l]_i=1$ belongs to $S_l$}{
						$\lm \leftarrow \lm + 1$\;
					}	
				} 
				\If{$\lm = L$} { 
					$\mathbf{\tilde{X}}_{b,k} \leftarrow min_{\mathbf{\phi}_{p}=1}{\{y_p\}}$\;
			 }	
		}
}
\caption{Filter($\mathbf{\Phi}, \mathbf{Y}_r, h$)}\label{algo:filter}
\end{algorithm} 

%
%
%
%

Next, the \textsc{Estimate} (Algorithm~\ref{algo:estimate}) algorithm determines
$S_r$ by computing a score for each BAC.  Based on
the computed scores, we select and return the top $s$ BACs as the
final support $S_r$ of $\mathbf{X}_r$. Read $r$ is then assigned to all the
BACs in $S_r$. The scoring function we employ for each BAC $b$ is the number of $k$-mers ``voting"
for $b$, \emph{i.e.}, having a frequency of at least $\tau$ in each
pool in the signature of $b$. The value we used for $\tau$ is
given in Section~\ref{sec:experiments}. If we consider the
rows of $\mathbf{\tilde{X}}_r$ as vectors of length $K_r$, our scoring
function is simply the $\l_0$ norm of these vectors, after zeroing out all the entries smaller than $\tau$. We also tried
$\l_1$ and $\l_2$ norms without observing significant improvements in the accuracy of read assignments. 

\begin{algorithm}[b]
	\DontPrintSemicolon 
	\SetFuncSty{textsc}
	\SetAlCapNameFnt{\sc}
	\SetKwData{Score}{score}
	\SetKwInOut{Input}{Input}\SetKwInOut{Output}{Output}

	\Input {$\mathbf{\tilde{X}}_r$, sparsity parameter $s$}
	\Output {Support set $S_r$, with $|S_r| \leq s$}

	\BlankLine

	\For{$b=1, \dots, n$}{
		$\Score{b} \leftarrow |\{k : \mathbf{\tilde{X}}_{b,k} \geq \tau\}|$ \;
	}

	$S_r \leftarrow$ set of indices $b$ with the highest $s$ scores \;
	
	\Return $S_r$

	\caption{Estimate$(\mathbf{\tilde{X}}_r, s)$}\label{algo:estimate}
\end{algorithm}

Observe that algorithms \textsc{FindSupport},
\textsc{Filter} and \textsc{Estimate} process one read at a
time. Since there is no dependency between the reads, processing
multiple reads in parallel is trivial. However, better total running
time, improved decoding accuracy as well as a smaller number of non-decodable reads can be achieved by
jointly decoding multiple reads at once. The idea is to use
additional sources of information about the reads, namely (1) read
overlaps and (2) mate-pair information. For the former, if we can determine
clusters of reads that are mutually overlapping, we can then decode
all the reads within a cluster as a single unit. Not only this strategy
increases the decoding speed, but it also has the potential to improve
the accuracy of read assignments because while some of
the reads in the cluster might have sequencing
errors, the others might be able to `compensate'. Thus, we can have more confidence
in the vote of high-quality shared $k$-mers. There is, however, the possibility that 
overlaps are misleading. For instance, overlaps between
repetitive reads might lead one to assign them to the same cluster
while in reality these reads belong to different BACs.  To
reduce the impact of this issue we allow any read that belongs to
multiple clusters to be decoded multiple times and take the
intersection of the multiple assignments as the final
assignment for the read. If a read does not overlap any other read
(which could be explained due to the presence of several sequencing
errors) we revert to the single read decoding strategy.
In order to build the clusters we compute all pairwise read overlaps
using \textsc{SGA} \cite{SGA}, whose parameters are discussed in
Section~\ref{sec:experiments}.


In order to apply \textsc{FindSupport} on a cluster $c$ of reads, we
need to gather the frequency matrix $\mathbf{Y}_c$ for $c$. Since the
total number of $k$-mers within a cluster can be quite large as
the clusters themselves can be quite large, and each $k$-mer can
be shared by a subset of the reads in the cluster, we build $\mathbf{Y}_c$
on the most frequently shared valid $k$-mers in the cluster.  Our
experiments indicate that retaining a number of $k$-mers
equal to the numbers of $k$-mers used in the decoding of
individual reads is sufficient.  When reads within a cluster do
not share a sufficient number of valid $k$-mers, we break the cluster
into singletons and decode its reads individually.  We denote by $\mu$ the
minimum number of valid $k$-mers required to attempt decoding of
both clusters and individual reads. The choice of this parameter is also discussed in
Section~\ref{sec:experiments}.

We can also use mate pair information to improve the decoding, if reads
are sequenced as paired-ends (PE). The \emph{mate resolution strategy} (MRS)
we employ is straightforward. Given a PE read $r$,
(1) if the assignment of one of the mates of $r$ is empty,
we assign $r$ to the BACs of the non-empty mate;
(2) if both mates of $r$ have BAC assignments and the intersection of these assignments is
non-empty, we assign $r$ to the BACs in the intersection; (3) if
both mates of $r$ have BAC assignments and their intersection is empty, we discard both mates.  In what follows, we will use \emph{RBD}
to refer to the read based-decoding and \emph{CBD} to refer to the cluster-based
decoding versions of our algorithm. CBD with MRS is summarized in Algorithm~\ref{algo:cluster}.

\begin{algorithm}[ht]
\DontPrintSemicolon 
\SetFuncSty{textsc}
\SetAlCapNameFnt{\sc}
\SetKwData{Read}{r}\SetKwData{Set}{S}
\SetKwFunction{FindSupport}{FindSupport}	
\SetKwInOut{Input}{Input}\SetKwInOut{Output}{Output}

\Input {$\mathbf{\Phi} \in \{0, 1\}^{m \times n}$, parameter h, sparsity parameter $s$, set $\mathcal{C}$ of all clusters, frequency matrix $\mathbf{Y}_c$ for each cluster $c \in \mathcal{C}$}
\Output {A support set $\Set_{\Read}$ with $|\Set_{\Read}|  \leq s$ for each read $\Read$}

\BlankLine

\For{each cluster $c \in \mathcal{C}$}{
	$\Set_c$ $\leftarrow $\FindSupport{$\mathbf{\Phi}, \mathbf{Y}_c,h,s$}\;
	
	\For{each read $\Read \in c$}{
		\lIf{$\Set_{\Read} = \emptyset$}{
			$\Set_{\Read} \leftarrow \Set_c$ \;
		} \lElse {
			$\Set_{\Read} \leftarrow \Set_{\Read} \cap \Set_c$  \tcp*[h] {Take intersection of all assignments to $r$}\;
		}	
	}
}
\tcp*[h]{MRS}\; 
\For{each PE read $(\Read_1, \Read_2)$} {
	\lIf{$\Set_{\Read_1} = \emptyset $}{
		$\Set_{\Read_1} \leftarrow \Set_{\Read_2}$\;
	} 
	\lIf{$\Set_{\Read_2} = \emptyset $}{
		$\Set_{\Read_2} \leftarrow \Set_{\Read_1}$\;
	} 
	 
	\If {$\Set_{\Read_1} \neq \emptyset $ and $\Set_{\Read_2} \neq \emptyset $} { 
	
		$\Set_{\Read_1,\Read_2} \leftarrow \Set_{\Read_1} \cap \Set_{\Read_2}$\;
		\If{$\Set_{\Read_1,\Read_2} \ne \emptyset$}{
			$\Set_{\Read_1} \leftarrow \Set_{\Read_1,\Read_2}$\;
			$\Set_{\Read_2} \leftarrow \Set_{\Read_1,\Read_2}$\;
		}
	}
}

\caption{ClusterFindSupport$(\mathbf{\Phi}, \mathcal{C}, \{\mathbf{Y}_c\}_{c \in \mathcal{C}},h,s)$}\label{algo:cluster}
\end{algorithm} 


\section{Experimental Results}
\label{sec:experiments}


While our algorithms can be used to decode any set of DNA samples pooled according to STD,
in this paper, we evaluate their performance on sets of BAC clones
selected in such a way that they cover the genome (or a portion thereof)
with minimum redundancy. In other words, the BACs we use form a
\emph{minimum tiling path} (MTP) of the genome.  The
construction of a MTP for a given genome requires a
physical map, but both
are well-known procedures and we will not discuss them here (see, \emph{e.g.},
\cite{FPC-MTP} and references therein).  Once the
set of MTP BAC clones has been identified, we (1) pool them
according to STD, (2) sequence the resulting pools, (3)
apply our decoding algorithm to assign reads
back to their source BACs. Step (3) makes it possible to assemble reads
BAC-by-BAC, thus simplifying the genome assembly
problem and increasing the accuracy
of the resulting BAC assemblies \cite{10.1371/journal.pcbi.1003010}. 

\begin{figure}[b]
  \centerline{\includegraphics{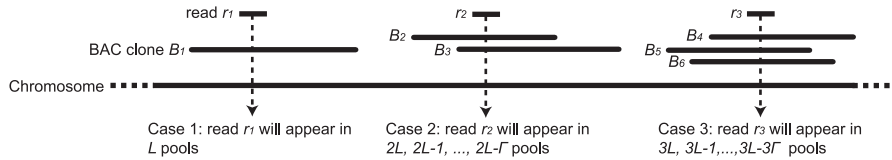}}
  \caption{The three cases we are dealing with during read decoding.}
  \label{fig:decoding_cases}
\end{figure}

Recall that CS decoding requires the unknown assignment vector
$\mathbf{x}$ to be $s$-sparse. Since we use MTP BAC clones, if the MTP
was truly a set of minimally overlapping clones, setting $s$ equal to
$2$ would be sufficient; we set it equal to $3$ instead to account for
imperfections in the construction of the MTP and to obtain additional
protection against errors. 
Figure~\ref{fig:decoding_cases} illustrates
the three cases (read belongs to one BAC, two BACs or three BACs) we
will be dealing with during decoding, and how it affects our STD parameter choice.

Next, we present experimental evaluations where we pool BAC clones using the
following STD parameters. Taking into consideration the need for a
$3$-decodable pooling design for MTP BACs, we choose parameters
$q=13$, $L=7$ and $\Gamma=2$, so that $m=qL=91$, $n=q^{\Gamma+1}=2197$
and $d=\lfloor{(L-1)/\Gamma}\rfloor=3$. In words, we pool $2197$ BACs
in $91$ pools distributed in $7$ layers of $13$ pools each. Each BAC
is pooled in exactly $7$ pools and each pool contains $q^\Gamma=169$
BACs. Recall that we call the set of $L$ pools to which a BAC is assigned the BAC signature. In the case of STD, any two-BAC signatures
can share at most $\Gamma=2$ pools and any three-BAC signatures
can share at most $3\Gamma=6$ pools. 


\subsection{Simulation results on the rice genome}\label{sec:rice}

To simulate our combinatorial pooling protocol and subsequent decoding, we used the genome of
rice (\emph{Oryza sativa}) which is about 390~Mb and fully
sequenced. We started from an MTP of 3,827 BAC clones selected from a
real physical map library for rice of 22,474 clones. The average BAC
length in the MTP was $\approx 150$kB. Overall the clones in the MTP spanned
91\% of the rice genome.  We pooled a subset of 2,197 of these BACs
into 91 pools according to the pooling parameters defined above.  The
resulting pools were `sequenced' \emph{in silico} using
\textsc{SimSeq}, which is a high-quality short read simulator used to
generate the synthetic data for Assemblathon
\cite{Earl:2011gt}. \textsc{SimSeq} uses error profiles derived from
real Illumina data to inject ``realistic'' substitution errors. For
each pool, we generated $10^6$ PE reads of $100$ bases each
with an average insert size of $300$ bases.  A total of $200$M usable
bases gave an expected $\approx 8\times$ sequencing depth for a BAC in
a pool. As each BAC is present in $7$ pools, this is an expected
$\approx 56\times$ combined coverage before decoding. After decoding
however, since a read can be assigned to more than one BAC,
the actual average BAC sequencing depth became $91.68\times$ for RBD, $93\times$
for CBD and $97.91\times$ for CBD with MRS.

To simulate our current workflow, we first performed
error-correction on the synthetic reads using \textsc{SGA}
\cite{SGA} with $k$-mer size parameter $k=26$. Then, the hash table for $k=26$ was built on the corrected
reads, but we only stored $k$-mers appearing in at least $\gamma=3$
pools.  Due to the error-correction preprocessing step and the fact
that we are discarding $k$-mers with low pool count, the hash table was
relatively small (about $30$GB).  

In order to objectively evaluate and compare the performance of our decoding algorithms, 
we first had to precisely define the `ground truth' for simulated reads. An easy
choice would have been to consider `true' only the single BAC from which each read was
generated. However, this notion of ground truth is not satisfactory: for instance,
since we can have two or three BACs overlapping
each other in the MTP, reads originating from an overlap region are expected to be assigned
to all the BACs involved. In order to find all the BACs that
contain a read, we mapped all synthetic reads (error-free version) against the BAC primary sequences using
\textsc{Bowtie} \cite{BOWTIE} in stringent mode (paired-end end-to-end
alignment with zero mismatches). The top three paired-end hits returned by \textsc{Bowtie} constituted
the ground truth against which we validated the accuracy of the
decoding.

In our experiments we observed that although the majority of the reads are
assigned to 1--3 BACs, due to the repetitive nature of the genome, a small
fraction ($\approx 1\%$) can be correctly assigned to more than $3$ BACs.
To account for this, rather than sorting BAC scores and retaining the top $3$,
we decided to assign a read to all BACs whose score was above a certain threshold.
We found that retaining all BACs whose score was at least $0.5 K_r$ gave the
best results. Recall that the score function we are using is the $\l_0$ norm,
so we are effectively asking that at least half of the $k$-mers `vote' for
a BAC. 


Table~\ref{tab:accuracy} summarizes and compares the decoding performance of our algorithms.
The first row of the table reports the performance of an `ideal'
method that always assigns each read to its original source BAC. The next four
rows summarize (1) the performance of \textsc{HashFilter} \cite{10.1371/journal.pcbi.1003010} with default parameters;
(2) our read-based decoding (RBD); (3) our cluster-based decoding (CBD); (4) our cluster-based decoding with mate
resolution strategy (CBD + MRS). For all three versions of the decoding algorithm we used parameters $h=\lfloor q/2 \rfloor =6$ and $\tau=1$. 

To build clusters, we require a minimum overlap of $75$ bases between two reads and a maximum error rate of $0.01$ (\textsc{SGA} parameters). The resulting clusters contained on average about $5$ reads. Our methods make a decoding decision if a read (or cluster) contains at least $\mu=15$ valid $k$-mers.
The columns in Table~\ref{tab:accuracy} report the percentage of reads assigned to the original source BAC,
\emph{precision} (defined as $TP/(TP+FP)$ where $TP$ is the number of 
true positive BACs across all decoded reads; $FP$ and $FN$ are
computed similarly), \emph{recall} (defined as $TP/(TP+FN)$),
\emph{F-score} (harmonic mean of precision and recall) and the percentage of
reads that were not decoded. Observe that the highest precision is achieved
by the cluster-based decoding with MRS, and the highest recall is obtained by 
\textsc{Hashfilter}. In general, all methods are comparable from the point
of view of decoding precision and recall. In terms of decoding time, once the hash table
is built ($\approx10$h on one core), RBD takes on average $14.03$s per $1$M reads and CBD takes on average $33.46$s per $1$M clusters. By comparison, \textsc{Hashfilter} \cite{10.1371/journal.pcbi.1003010} takes about $30$s per 1M reads. These measurements were done on $10$ cores of an Intel Xeon X5660 2.8~GHz server with 12 cores and 192 GB of RAM. 

\begin{table*}[t]
\begin{center}
\begin{tabular}{|l|r|r|r|r|r|}
\hline
                & \emph{Mapped to source BAC}  & \emph{Precision} & \emph{Recall} & \emph{F-score} & \emph{Not decoded}  \\ 
\hline\hline
Perfect decoding        & 100.00\% & 98.11\% & 49.62\% & 65.90\% & 0.00\% \\  \hline \hline
\textsc{Hashfilter} \cite{10.1371/journal.pcbi.1003010} & \textbf{99.48\%} & 97.45\% & \textbf{99.28\%} & \textbf{98.36\%} & 16.25\% \\ \hline  
RBD             & 98.05\% & 97.81\% & 97.46\% & 97.64\% & 14.58\% \\  \hline
CBD          & 97.23\% & 97.74\% & 96.35\% & 97.04\% & 12.58\%  \\ \hline  
CBD + MRS & 96.60\% & \textbf{97.89\%} & 95.58\% & 96.72\% & \textbf{7.09\%} \\ \hline 
\end{tabular} 
\end{center}
\caption{Accuracy of the decoding algorithms on synthetic reads for the rice genome (see text for details). All values are an average of 91 pools. Boldface values highlight the best result in each column (excluding perfect decoding).} 
\label{tab:accuracy}
\end{table*}

As a more meaningful measure of decoding performance, we
assembled the set of reads assigned by each method to each BAC.  We carried out this
step using \textsc{Velvet} \cite{Velvet08} for each of the 2,197 BACs, using a
range of $l$-mer from $25$ to $79$ with an increment of $6$, and chose
the assembly that achieved the highest N50\footnote{The N50 is the contig length such that at least half of the total  bases of a genome assembly are contained within contigs of this length or longer.}. Table~\ref{tab:rice_assembly} reports the main statistics for the assemblies:
percentage of reads used by \textsc{Velvet} in the assembly, number of contigs
(at least $200$ bases long) of the assembly, value of N50, ratio of the sum of all contigs
sizes over BAC length, and the coverage of the BAC primary sequence by the assembly. All reported values are averages over 2,197 BACs.
We observe that our decoding algorithms lead to superior assemblies than \textsc{Hashfilter}'s. In particular,
the N50 and the average coverage of the original BACs are both very
high, and compare favorably with the statistics for the assembly of perfectly decoded
reads. 

The discrepancy between similar precision/recall figures but quite different assembly statistics deserves  a comment. First, we acknowledge that the way we compute precision and recall by averaging $TP$, $FP$ and $FN$ across all decoded reads might not be the best way of measuring the accuracy of the decoding. Taking averages might not accurately reflect mis-assignments at the level of individual reads. Second, our  decoding algorithms makes a better use of the $k$-mer frequency information than \textsc{HashFilter}, and, at the same time, takes advantage of overlap and mate pair information, which is expected to result in more reads decoded and more accurate assemblies.

\begin{table*}[t]
\begin{center}
\begin{tabular}{|l|r|r|r|r|r|r|r|r|r|}
\hline
                                & \emph{Reads used} & \emph{\# of contigs} & \emph{N50} & \emph{Sum/size} & \emph{BAC coverage} \\ 
\hline\hline
Perfect decoding (ideal)        & 97.1\% &  4 & 136,570 & 107.4 & 87.1\% \\ \hline \hline
\textsc{Hashfilter} \cite{10.1371/journal.pcbi.1003010} & 95.0\% & 24 &  52,938 &  93.8 & 76.2\% \\ \hline  
RBD             & 96.5\% & 20 &  46,477 &  90.0 & 81.1\% \\ \hline
CBD          & \textbf{97.3\%} & 22 &  53,097 &  93.8 & \textbf{84.7\%} \\ \hline  
CBD + MRS & 97.0\% & \textbf{11} & \textbf{103,049} &  97.0 & 82.9\% \\ 
\hline 
\end{tabular} 
\end{center}
\caption{Assembly results for rice BACs for different decoding algorithms (see text for details). All values are an average of 2197 BACs. Boldface values highlight the best result in each column (excluding perfect decoding).}
\label{tab:rice_assembly}
\end{table*}


\subsection{Results on the barley genome} 

We have also collected experimental results on real sequencing data for
the genome of barley (\emph{Hordeum vulgare}), which is about 5,300~Mb and 
at least $95\%$ repetitive. We started from an MTP of about
15,000 BAC clones selected from a subset of nearly 84,000 gene-enriched
BACs for barley (see \cite{10.1371/journal.pcbi.1003010} for more details).
We divided the set of MTP BACs into seven sets of $n=2197$ BACs and pooled each
set using the STD parameters defined above. In this manuscript, we
report on one of these seven sets, called HV3 (the average BAC length in this set
is about $116$K bases). The $91$ pools in HV3 were sequenced on one flow cell of
the Illumina HiSeq2000 by multiplexing $13$ pools on each lane. After each
sample was demultiplexed, we quality-trimmed and cleaned the reads of spurious
sequencing adapters and vectors. We ended up with high quality reads of about
$87$--$89$ bases on average. The number of reads in a pool
ranged from $4.2$M to $10$M, for a grand total of $826$M reads.   
We error-corrected and overlap-clustered the reads using \textsc{SGA} (same parameters as for rice). The average cluster size was about $26$ reads. Computing pairwise
overlaps took an average of $217.60$s per 1M reads on $10$ cores.
The hash table for $k=26$ (after discarding $k$-mers appearing in fewer
than $\gamma=3$ pools) used about $26$GB of RAM.     
After decoding the reads to their BAC, we obtained an average
sequencing depth for one BAC of $409.2\times$, $382.2\times$ and
$412.8\times$ for RBD, CBD and CBD + MRS, respectively. The average running
time was $10.25$s per 1M reads for RBD and $82.12$s per 1M clusters for
CBD using $10$ cores.   

The only objective criterion to asses the decoding performance on barley genome
is to assemble the reads BAC-by-BAC and analyze the assembly statistics. We used
\textsc{Velvet} with the same $l$-mer choices as used for rice. Table~\ref{tab:barley_assembly}
summarizes the statistics for the highest N50 among those $l$-mer choices.
As before, rows corresponds to the various decoding methods. 
Columns show (1) percentage of reads used by \textsc{Velvet} in the assembly, (2)
number of contigs (at least $200$ bases long), (3) value of N50, (4)
ratio of the sum of all contigs sizes over estimated BAC length, (5)
the number of barley known unigenes observed in the assemblies, and (6) the coverage of
observed unigenes. Observe that, out of a total of 1,471 known unigenes 
expected to be contained in these BACs, a large fraction are reported
by all assemblies. However, cluster-based decoding appears to generate
significantly longer contigs than the other methods.

\begin{table*}[t]
\small
\begin{center}
\begin{tabular}{|l|r|r|r|r|r|r|}
\hline
           & \emph{Reads used} & \emph{\# contigs} & \emph{N50} & \emph{Sum/size} & \emph{\# obs unigenes} & \emph{\% coverage} \\ \hline \hline
\textsc{Hashfilter} \cite{10.1371/journal.pcbi.1003010} & 83.6\% & 101 & 8,190 & 96.7\% & 1,433 & 92.9\% \\ \hline
RBD           & 85.7\% & 54 & 14,419 & 101.0\% & 1,434 & 92.4\%  \\ \hline
CBD        & 92.9\% & 54 & 13,482 & 94.5\% & \textbf{1,436} & \textbf{92.6\%}\\ \hline
CBD + MRS  & \textbf{94.3\%} & \textbf{50} & \textbf{26,842} & 126.8\% & 1,434 & 92.5\%  \\ \hline 
\end{tabular} 
\end{center}
\caption{Assembly results for barley BACs for different decoding algorithms. All values are an average of 2197 BACs. Boldface values highlight the best result in each column. Column ``\% coverage'' refers to the coverage of known unigenes by assembled contigs.} 
\label{tab:barley_assembly}
\end{table*}


\section{Conclusions}

We have presented a novel modeling and decoding approach for pooled sequenced reads obtained from 
protocols for \emph{de novo} genome sequencing, like the one
proposed in \cite{10.1371/journal.pcbi.1003010}. Our algorithm is based on the theory of compressed sensing and uses 
ideas from the decoding of error-correcting codes. It also effectively exploits overlap and mate pair information between the sequencing reads. Experimental results on synthetic
data from the rice genome as well as real data from the genome of barley show that
our method enables significantly higher quality
assemblies than the previous approach, without incurring higher decoding
times. 


\section*{Acknowledgments}

SL and TJC were supported by NSF [DBI-1062301 and DBI-0321756] and by USDA [2009-65300-05645 and 2006-55606-16722]. MW and ACG were supported by NSF [CCF-1161233]. HQN and AR were supported by NSF [CCF-1161196].



\begin{thebibliography}{10}

\bibitem{BarcodeBias}
S.~Alon, F.~Vigneault, S.~Eminaga, \emph{et al.}
\newblock Barcoding bias in high-throughput multiplex sequencing of mirna.
\newblock {\em Genome Research}, 21(9):1506--1511, 2011.

\bibitem{Amir&Zuk}
A.~Amir and O.~Zuk.
\newblock Bacterial community reconstruction using compressed sensing.
\newblock In {\em RECOMB}, pages 1--15, 2011.

\bibitem{Earl:2011gt}
D.~Earl and \emph{et al.}
\newblock {Assemblathon 1: A competitive assessment of de novo short read
  assembly methods}.
\newblock {\em Genome Research}, 21(12):2224--2241, Dec. 2011.

\bibitem{FPC-MTP}
F.~W. Engler, J.~Hatfield, W.~Nelson, and C.~A. Soderlund.
\newblock Locating sequence on {FPC} maps and selecting a minimal tiling path.
\newblock {\em Genome Research}, 13(9):2152--2163, 2003.

\bibitem{YanivErlich072009}
Y.~Erlich, K.~Chang, A.~Gordon, \emph{et al.}
\newblock {DNA} sudoku -- harnessing high-throughput sequencing for multiplexed
  specimen analysis.
\newblock {\em Genome Research}, 19(7):1243--1253, 2009.

\bibitem{Erlich:2010je}
Y.~Erlich, A.~Gordon, M.~Brand, \emph{et al.}
\newblock Compressed genotyping.
\newblock {\em IEEE Transactions on Information Theory}, 56(2):706--723, Apr.
  2010.

\bibitem{Hajirasouliha:2008}
I.~Hajirasouliha, F.~Hormozdiari, S.~C. Sahinalp, and I.~Birol.
\newblock Optimal pooling for genome re-sequencing with ultra-high-throughput
  short-read technologies.
\newblock {\em Bioinformatics}, 24(13):i32--i40, 2008.

\bibitem{BOWTIE}
B.~Langmead, C.~Trapnell, M.~Pop, and S.~L. Salzberg.
\newblock Ultrafast and memory-efficient alignment of short {DNA} sequences to
  the human genome.
\newblock {\em Genome Biology}, 10(3):R25, 2009.

\bibitem{10.1371/journal.pcbi.1003010}
S.~Lonardi, D.~Duma, M.~Alpert, \emph{et al.}
\newblock Combinatorial pooling enables selective sequencing of the barley gene
  space.
\newblock {\em PLoS Comput Biol}, 9(4):e1003010, 04 2013.

\bibitem{NgoPoratRudra12}
H.~Q. Ngo, E.~Porat, and A.~Rudra.
\newblock Efficiently decodable compressed sensing by list-recoverable codes
  and recursion.
\newblock In {\em STACS}, pages 230--241, 2012.

\bibitem{SnehitPrabhu072009}
S.~Prabhu and I.~Pe'er.
\newblock Overlapping pools for high-throughput targeted resequencing.
\newblock {\em Genome Research}, 19(7):1254--1261, 2009.

\bibitem{Shental:2010bu}
N.~Shental, A.~Amir, and O.~Zuk.
\newblock Identification of rare alleles and their carriers using compressed
  se(que)nsing.
\newblock {\em Nucleic Acids Research}, 38(19):e179--e179, 2010.

\bibitem{SGA}
J.~T. Simpson and R.~Durbin.
\newblock Efficient de novo assembly of large genomes using compressed data
  structures.
\newblock {\em Genome Research}, 22(3):549--556, 2012.

\bibitem{Nature12}
{The International Barley Genome Sequencing Consortium}.
\newblock A physical, genetic and functional sequence assembly of the barley
  genome.
\newblock {\em Nature}, advance online publication:in press, 10 2012.

\bibitem{Thierry2006a}
N.~Thierry-Mieg.
\newblock A new pooling strategy for high-throughput screening: the shifted
  transversal design.
\newblock {\em BMC Bioinformatics}, 7(28), 2006.

\bibitem{OMP}
J.~A. Tropp and A.~C. Gilbert.
\newblock Signal recovery from random measurements via orthogonal matching
  pursuit.
\newblock {\em IEEE Trans. Inform. Theory}, 53:4655--4666, 2007.

\bibitem{SOMP}
J.~A. Tropp, A.~C. Gilbert, and M.~J. Strauss.
\newblock Algorithms for simultaneous sparse approximation: part i: Greedy
  pursuit.
\newblock {\em Signal Process.}, 86(3):572--588, 2006.

\bibitem{Velvet08}
D.~Zerbino and E.~Birney.
\newblock {V}elvet: {A}lgorithms for \emph{de novo} short read assembly using
  de {B}ruijn graphs.
\newblock {\em Genome Research}, 8(5):821--9, 2008.

\end{thebibliography}
\end{document}